\documentclass[10pt,conference]{IEEEtran}
\IEEEoverridecommandlockouts
\usepackage{cite}
\usepackage{amsmath,amssymb,amsfonts}
\usepackage{graphicx}
\usepackage{textcomp}
\usepackage{booktabs} 
\usepackage{todonotes}
\usepackage{graphicx}
\usepackage{tikz}
\usepackage{amsmath} 
\usepackage[inline]{enumitem}
\usetikzlibrary{arrows}
\usepackage{lipsum}
\usepackage{makecell}
\usepackage{fmtcount} 
\usepackage{array,multirow}
\usepackage{xspace}
\usepackage{url}
\usepackage{amsmath,scalerel}
\usepackage{fancyvrb}
\usepackage{xcolor}
\usepackage{moreverb}
\usepackage{listings}    
\usepackage{alltt,xcolor}

\newcommand*{\LargerCdot}{\raisebox{-1ex}{\scalebox{3}{$\cdot$}}}
\newcommand{\etal}{et al.\xspace}
\newcommand{\ie}{i.e.,\xspace}
\newcommand{\eg}{e.g.,\xspace}
\newcommand{\fig}[1]{Figure~\ref{#1}}
\newcommand{\tab}[1]{Table~\ref{#1}}

\newcommand{\github}{\emph{GitHub}\xspace}

\newcommand{\skopeo}{\emph{skopeo}\xspace}

\newcommand{\linux}{\emph{Linux}\xspace}
\newcommand{\docker}{\emph{Docker}\xspace}
\newcommand{\official}{\emph{official}\xspace}
\newcommand{\community}{\emph{community}\xspace}
\newcommand{\debian}{\emph{Debian}\xspace}

\newcommand{\stretchh}{\emph{Stretch}\xspace}
\newcommand{\jessie}{\emph{Jessie}\xspace}
\newcommand{\buster}{\emph{Buster}\xspace}

\newcommand{\udd}{\emph{UDD}\xspace}

\newcommand{\alpine}{\emph{Alpine}\xspace}
\newcommand{\dockerhub}{\emph{Docker Hub}\xspace}

\def\BibTeX{{\rm B\kern-.05em{\sc i\kern-.025em b}\kern-.08em
    T\kern-.1667em\lower.7ex\hbox{E}\kern-.125emX}}

\begin{document}

	\title{On The Relation Between Outdated Docker Containers, Severity Vulnerabilities and Bugs}


	\author{\IEEEauthorblockN{Ahmed Zerouali}
       \IEEEauthorblockA{ahmed.zerouali@umons.ac.be}
       \IEEEauthorblockA{University of Mons}
		\and
		\IEEEauthorblockN{Tom Mens}
		\IEEEauthorblockA{tom.mens@umons.ac.be}
      \IEEEauthorblockA{University of Mons}
		\and
		\IEEEauthorblockN{Gregorio Robles }
		\IEEEauthorblockA{grex@gsyc.urjc.es}
      \IEEEauthorblockA{Universidad Rey Juan Carlos}
		\and
		\IEEEauthorblockN{Jesus M. Gonzalez-Barahona}
		\IEEEauthorblockA{jgb@gsyc.es}
      \IEEEauthorblockA{Universidad Rey Juan Carlos}
 
}

\maketitle

\begin{abstract}
Packaging software into containers is becoming a common practice when deploying services in cloud and other environments. \docker images are one of the most popular container technologies for building and deploying containers. A container image usually includes a collection of software packages, that can have bugs and security vulnerabilities that affect the container health. 
Our goal is to support container deployers by analysing the relation between \textit{outdated} containers and \textit{vulnerable} and \textit{buggy} packages installed in them. 
We use the concept of technical lag of a container as the difference between a given container and the most up-to-date container that is possible with the most recent releases of the same collection of packages. For 7,380 \official and \community \docker images that are based on the \debian \linux distribution, we identify which software packages are installed in them and measure their technical lag in terms of version updates, security vulnerabilities and bugs. 
We have found, among others, that no release is devoid of vulnerabilities, so deployers cannot avoid vulnerabilities even if they deploy the most recent packages.
We offer some lessons learned for container developers in regard to the strategies they can follow to minimize the number of vulnerabilities.
We argue that \docker container scan and security management tools should improve their platforms by adding data about other kinds of bugs and include the measurement of technical lag to offer deployers information of when to update.
\end{abstract}

\begin{IEEEkeywords}
Empirical analysis, \docker containers, technical lag, security vulnerability\end{IEEEkeywords}

\section{Introduction}
\label{sec:intro}

Packaging software into containers has become a common practice during the last years~\cite{bernstein2014containers}. In particular, \docker containers are a popular schema to provision multiple software applications on a single host. A container is a running image, which includes its own system libraries, configuration files, and software~\cite{merkel2014docker}, providing support for both \linux-based and other operating systems~\cite{turnbull2014docker,mouat2015using}. \docker allows for the creation of registries, providing a common place to share \docker images.
With more than $1.6M$ images (October 2018), \dockerhub\cite{dockerhub} is one of the largest of such registries.



Images in \dockerhub are organized in \textit{repositories}, each one providing a set of versioned \docker images. Repositories can be private or public, which in turn are split into  \official and \community repositories.  An official repository contains public and certified images from recognized vendors (\eg ElasticSearch, Debian, Alpine). Images in official repositories are frequently used as the base for other \docker images, since they are supposed to be secure and well maintained. Community repositories can be created by any user or organization~\cite{boettiger2015introduction}.  


When \docker images are built with \linux-based operating systems, they usually follow the packaging model for their \linux distribution of choice, with most of the software they include installed as a package. Once the image is built, packages remain \textit{frozen} (for a certain version of that image). From time to time, a new version of the image is built, with a newer version of the packages. But the old version may be still in use, deployed as a container in production. Those containers corresponding to old images may include outdated packages with known security vulnerabilities and bugs, already fixed in newer versions but still present in them. Since the containers may run in production, they could be exposed to exploits of those vulnerabilities, and problems due to those bugs.

On the other hand, deployers of containers may prefer to stick to old versions, because they are known to work well and have been tested in production for a long time. In fact, the use of container images provides some isolation from evolving dependencies and changes in packages that may break working systems. This is a strong motivation to stick to an image which ``just works'', even if it is outdated, since upgrading to new versions of container images always involves some risk.
Thus, deployers are always balancing their need to update to new images --with fixed vulnerabilities and bugs-- and the risk of breaking a working system with the upgrade, due to unexpected changes in the packages.


This compromise has been widely reported in literature. According to a survey carried out in January 2015 by Red Hat and Forrester~\cite{dockerSurvey15}, security is a top concern when deciding whether to deploy containers. Another survey carried out by \textit{DevOps.com} and \textit{RedMonk} in December 2016~\cite{dockerSurvey16} revealed that users who are more concerned by image security focused on scanning simple \textit{Common Vulnerabilities and Exposures} (CVE) on the operating system. 
In April 2017, \textit{Anchore.io} conducted a survey with 242 users on the current landscape of practices being deployed by container users~\cite{anchoreSurvey17}. 
One of the questions of that survey was: \textit{``Other than security, what are the other checks  that you perform before running application containers?"} The top answers related to software package were: required packages ($\sim40\%$ of the answers); presence of bugs in major third-party software ($\sim33\%$ of the answers); and verifying whether third party software versions are up-to-date ($\sim27\%$ of the answers).

To support deployers of containers in this everyday compromise, we propose a method to assess on how \textit{outdated}, \textit{vulnerable}, and \textit{buggy} \docker images are with respect to the latest available releases of the packages they include.

The method is based on the concept of \textit{technical lag}~\cite{gonzalez2017technical}, which we use to estimate the difference between the software deployed in production and the most recent version of this software (in our case, in terms of novelty, vulnerabilities and bugs).
To show the applicability of the approach, we have conducted an empirical study, measuring technical lag, security vulnerabilities and bugs for $2,453$ official and $4,927$ community \dockerhub images based on the \debian \linux distribution.
The research questions that we have addressed in this study are:

\noindent {\bf $RQ_0$}: How often are \docker images updated?

\noindent {\bf $RQ_1$}: What is the technical lag induced by outdated packages in containers?

\noindent {\bf $RQ_2$}: How vulnerable are packages in containers?

\noindent {\bf $RQ_3$}: To which extent do containers suffer from bugs in packages?

\noindent {\bf $RQ_4$}: How long do bugs remain unfixed?

\noindent {\bf $RQ_5$}: How long do security vulnerabilities remain unfixed?


The rest of this paper is structured as follows:
Section~\ref{sec:related} reviews some relevant related work;
Section~\ref{sec:method} outlines the empirical methodology, including the datasets used, the selection of the analyzed images, and the collection of data;
Section~\ref{sec:results} presents the results of the empirical analysis;
Section~\ref{sec:discussion} highlights our contributions and main findings;
Section~\ref{sec:threats} discusses the limitations of the study;
and finally Section~\ref{sec:conclusion} outlines possible directions for future work, and conclusions.

\section{Related work}
\label{sec:related}

Gonz\'alez-Barahona \etal~\cite{gonzalez2017technical} proposed a theoretical model of ``technical lag'' to measure how outdated software components are. They explored many ways in which technical lag can be measured, and presented specific cases for which it is useful to analyze the evolution of technical lag.
 
Kula \etal~\cite{Kula2017:EMSE} studied the impact of dependency updates in the \github ecosystem. They empirically studied library migration of a set of 4,600 \github repositories and 2,700 library dependencies, and found that 81.5\% of the studied projects keep their outdated dependencies. Surveying developers about this, they found that 69\% of the interviewees were unaware of these outdated dependencies. Zerouali \etal\cite{zerouali2018empirical} introduced and analyzed a technical lag metric for dependencies in package networks, in order to assess how outdated a software package is compared to the latest available releases of its dependencies. Considering \textit{JavaScript} packages as a case study, they found a strong presence of technical lag caused by the specific use of dependency constraints. Decan \etal\cite{decan2018evolution} found similar results. Cox \etal \cite{cox2015measuring} analyzed 75 software systems and introduced different metrics to quantify their use of recent versions of dependencies. They found that systems using outdated dependencies were four times more likely to have security issues than up-to-date systems.

Focusing on \docker images, Cito \etal~\cite{cito2017empirical} conducted an empirical study on a dataset of 70,000 Dockerfiles, and contrasted this general population with samplings containing the top 100 and top 1,000 most popular projects using \docker. Their goal was to characterize the \docker ecosystem, discover prevalent quality issues, and study the evolution of Docker images. Among other results, they found that the most popular projects change more often than the rest of the \docker population, with an average of 5.81 revisions per year and 5 lines of code changed. Furthermore, they found that, from a representative sample of 560 projects, 34\% of all Docker images could not be built. 

Shu \etal~\cite{shu2017study} performed a generic large scale study on the state of security vulnerabilities in both community and official \dockerhub repositories. They proposed the \emph{Docker Image Vulnerability Analysis (DIVA)} framework to automatically discover, download, and analyze \docker images for security vulnerabilities. They studied a set of 356,218 images and observed that both official and community repositories contain on average more than 180 vulnerabilities; many images had not been updated for hundreds of days, demonstrating a strong need for more analysis and systematic methods of studying the content of \docker containers.

In our work, we only consider unique images, since we found many duplicate images inside the same repository. 
Additionally, we study in more detail the relation between package security vulnerabilities and technical lag of outdated packages.
Moreover, and to the best of our knowledge, we are the first to study and report results about non-security-related package bugs in \docker containers.

\section{Method and Data extraction}
\label{sec:method}

  
Our study is based on pulling \docker images from \dockerhub, identifying which packages are installed in them, and computing the technical lag for the image by aggregating the technical lag of those packages. We will measure the technical lag of individual packages in terms of version updates, vulnerabilities, and bugs. Our initial sample is composed of all official images in \dockerhub which are based on Debian, and the most pulled images based on \debian. Therefore, we only need to compute technical lag for \debian packages.

\begin{figure}[!ht]
	\begin{center}
		\setlength{\unitlength}{1pt}
		\footnotesize
		\includegraphics[width=0.8\columnwidth]{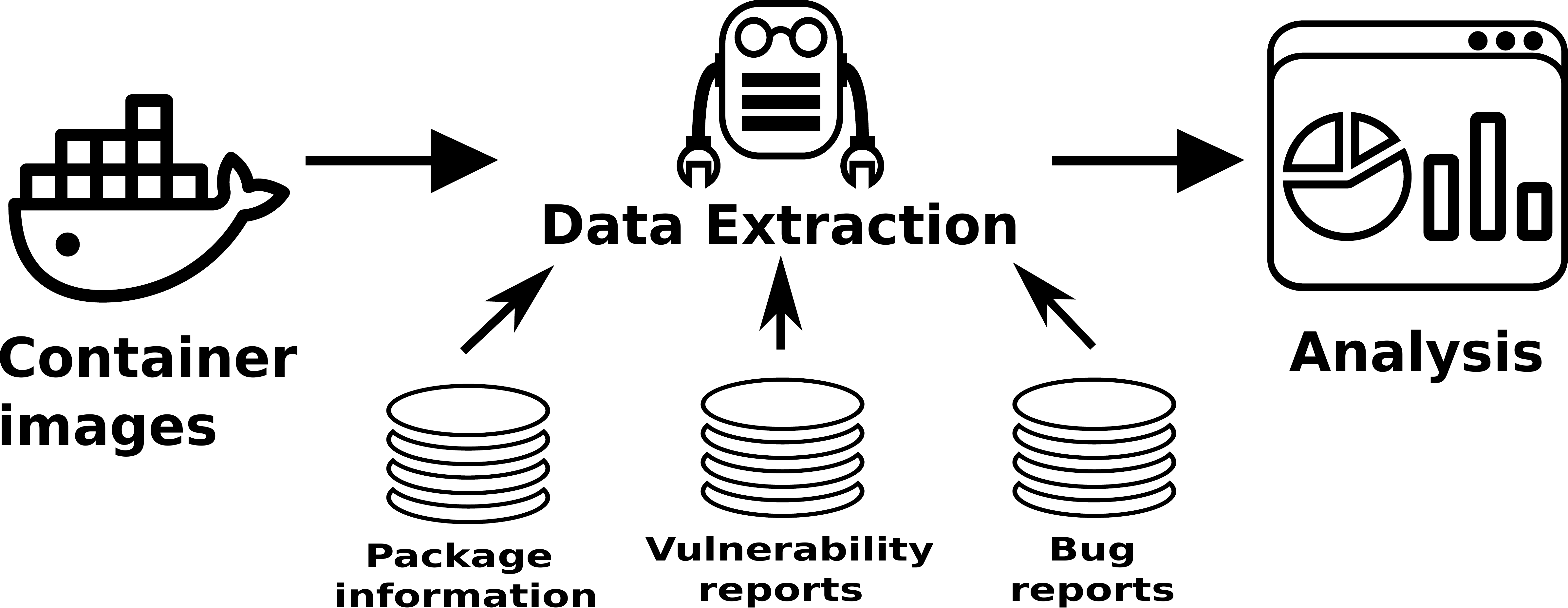}
		\caption{Process of the \docker container package analysis.}
		\label{fig:conpan}
	\end{center}

\end{figure}

The overall process, which we describe in detail below, is: (1) identification of \dockerhub base images for \debian, defining our base set; (2) identification of \dockerhub images in our dataset, including those derived from the base set; (3) analysis of all those images, matching their packages to a historical archive of all \debian packages; and (4) identification of bug and vulnerability reports for those packages, based on a historical database with those details for \debian packages.
\fig{fig:conpan} shows how we used the main data sources for our study. The next subsections explain in detail how we gathered the used datasets.


A replication package for our study is available for download at \url{https://doi.org/10.5281/zenodo.1250314}.

\subsection{Base images for \debian}

We decided to work with \docker images based on a \linux-distribution, because applications in them are usually installed using well-defined packages. Among them, we selected \debian because of its maturity and widespread use\footnote{https://www.ctl.io/developers/blog/post/docker-hub-top-10/} in \dockerhub. On October 1\textsuperscript{st} 2018, the \debian repository on \dockerhub had more than 125M pulls\footnote{\url{https://registry.hub.docker.com/v2/repositories/library/debian/}}.

While it is possible to create \docker images from scratch, most of them are based on others, which in the end are built on \textit{base images} that do not rely on any other image, except for the \docker-reserved minimal image named ``scratch''\footnote{https://docs.docker.com/develop/develop\-images/baseimages/}. Since we want to deal with images based in \debian, we first identified \debian base images\footnote{https://hub.docker.com/\_/debian/}.



The \debian project maintains packages for several simultaneous release lines (\debian distributions)~\cite{gonzalez2009macro}. The most important distributions are \textit{Testing, Stable} and \textit{Oldstable}. In \textit{Testing}, packages are updated frequently, when new releases have been inspected and validated (\eg lack of critical bugs, successful compilation, etc). At some points in time, when \textit{Testing} as a whole reaches a certain level of quality and stability, it is ``frozen'', and their packages used to produce a new \textit{Stable} distribution. Upon release of a \textit{Stable} version, the former one becomes \textit{Oldstable}, which in turn becomes \textit{Oldoldstable}. While updates in \textit{Testing} usually come with new functionality, updates in \textit{Stable} and \textit{Oldstable} include only the most important fixes or security updates. Currently, there is no security support for \textit{Oldoldstable} and older distributions. Thus, we chose to analyze \dockerhub \debian base images only for \textit{Testing, Stable} and \textit{Oldstable}. \tab{debian_versions} shows general information about the \debian versions considered for this work.
\begin{table}[!ht]
\centering
\caption{General information about the analyzed \debian versions.}
\label{debian_versions}
\begin{tabular}{c|c|c|c}
Version name    & Version  & Distribution     & Release date as stable\\ \hline
\buster  & \debian 10      & Testing   & -            \\ 
\stretchh & \debian 9       & Stable    & 2017-06-17   \\ 
\jessie  & \debian 8       & Oldstable & 2013-04-25   
\end{tabular}
\end{table}


\subsection{Identifying analyzed images}

Images in \dockerhub are named with the name of the repository, followed by a colon, and a tag (\textit{``imageRepo:Tag''}). Any image can be tagged more than once, and therefore may have more than one name (e.g., \textit{``debian:testing''}, \textit{``debian:testing-20181011''}). In the case of community images, the name of the repository usually starts with the name of the organization producing the images: \textit{``organizationName/ImageName''}. Therefore, full image names tend to have the form \textit{``organizationName/ImageName:Tag''}.


Each image is composed of one or many \textit{intermediate images} called \textit{layers}. Each layer is related to a change caused by commands that happened in the Dockerfile\footnote{https://docs.docker.com/engine/reference/builder/} used to produce the image, and has a unique hash signature. For example, the Dockerfile of the \textit{debian:stretch} image is:\\
\noindent\fbox{%
	\parbox{0.48\textwidth}{%
	\vspace{-0.1cm}
\begin{alltt}\small
\textcolor{red}{FROM} scratch

\textcolor{red}{ADD} rootfs.tar.xz /

\textcolor{red}{CMD} ["bash"]
\end{alltt}
\vspace{-0.1cm}
}}\\

When building the image with this Dockerfile, a single layer is produced:

\noindent\fbox{%
	\parbox{0.48\textwidth}{%
	\vspace{-0.1cm}
\begin{alltt}\small
\textcolor{blue}{debian:stretch Layers:} \textcolor{red}{[} "sha256:e1df5dc88d2cc2cd9a1b1680ec3cb" \textcolor{red}{]}
\end{alltt}
\vspace{-0.1cm}
}}\\

This image, in turn, can be used in other Dockerfiles as their base image using \textit{``FROM debian:stretch''}. Each image produced from those Dockerfiles will contain the layer(s) of the base image. For example, \textit{debian:stretch-backports} is produced with this Dockerfile:

\noindent\fbox{%
	\parbox{0.48\textwidth}{%
\begin{alltt}\small
\vspace{-0.1cm}
\textcolor{red}{FROM} debian:stretch

\textcolor{red}{RUN} echo 'deb http://deb.debian.org/debian
stretch-backports main'
 
> /etc/apt/sources.list.d/backports.list
\end{alltt}
\vspace{-0.1cm}
}}\\

The resulting \textit{debian:stretch-backports} image includes the layer found in \textit{debian:stretch}, its base image:

\noindent\fbox{%
	\parbox{0.48\textwidth}{%
	\vspace{-0.1cm}
\begin{alltt}\small
\textcolor{blue}{debian:stretch-backports Layers:} \textcolor{red}{[} "sha256:e1df5dc88d2cc2cd9a1b1680ec3cb"\textcolor{red}{,} 
"sha256:4c1b8d4c6076530dcd195cbc309e5" \textcolor{red}{]}
\end{alltt}
\vspace{-0.1cm}
}}\\

Therefore, we can identify \docker images derived from \debian base images by checking if they contain their layers.
Using the \dockerhub API we extracted all available image names from the 124 \official repositories, and those with at least 500 pulls from the top 30,000 \community repositories (by number of pulls). Using the \skopeo tool\footnote{\skopeo is a utility to inspect a repository on a Docker registry: \url{https://github.com/containers/skopeo}} we inspected images corresponding to all those image names, identifying unique images, and finding which ones included layers from our set of \debian base images. From 14,653 image names in \official repositories, we found 2,453 unique images (\ie 4,769 names) based on our set of \debian images. From 30,000 \community repositories, we found 4,927 unique images derived form our \debian set. All of them together composed our dataset of 7380 images.
\tab{images_debian} shows the number of images found for each \debian version.

\begin{table}[!ht]
\centering
\caption{Number of \docker images per \debian distribution.}
\label{images_debian}
\begin{tabular}{r|r|r|r}
Containers & \buster~/ Testing & \stretchh~/ Stable & \jessie~/ Oldstable \\ \hline
Official   & 150       & 620      & 1,683        \\ 
Community  & 86        & 1,248    & 3,593         \\ \hline
Total  & 236        & 1,868    & 5,276         \\ 
\end{tabular}
\end{table}


\subsection{Identifying installed packages}

\docker containers based in \debian include specific versions of \debian binary packages. Binary packages are produced from source packages, which we need to identify because we use them to find vulnerabilities and bug reports.

For tracking binary packages, and finding their metadata (including the name and version of the source package from which they were produced), we extracted daily snapshots of all \textit{amd64} binary packages for \textit{Oldstable} (\jessie), \textit{Stable} (\stretchh) and \textit{Testing} (\buster) distributions from the official and security \debian Snapshot repositories\footnote{http://snapshot.debian.org/archive/debian/ and  http://snapshot.debian.org/archive/debian-security/}.


Then, we pulled each \docker image in our images dataset, and identified its packages using regular \debian tools (\verb|dpkg -l|). We matched them to our dataset obtained from \debian Snapshot, finding in it more than 99\% of the packages in our images (1,379,163 package versions in \official images, and 561,982 in \community images).


\subsection{Vulnerability reports}

To find out known security vulnerabilities for the \debian packages in our dataset of \docker images, we used the \debian Security Bug Tracker\footnote{https://security-tracker.debian.org/tracker/data/json} as of 2018-03-18. For each package, the status of known vulnerabilities is maintained by the \debian Security Team, using data from different data sources (CVE database\footnote{https://cve.mitre.org/cve/}, National Vulnerability Database "NVD"\footnote{https://nvd.nist.gov}, etc).


In this \debian tracker, information about vulnerabilities is maintained at the source package level. A \debian vulnerability report contains information about affected source packages, severity, status, \debian bug id (if available), affected distributions, fixed version (if available), etc. Using it we can link vulnerabilities to source packages, and from there (using the \debian Snapshot dataset) to binary packages in our container images of interest. 
For each reported vulnerability for a package present in one of the analyzed images, we say that the corresponding package version is \emph{vulnerable} if the vulnerability is still open, or the vulnerability has been fixed in a more recent version than the one installed in the image.

\subsection{Bug reports}

For bug reports, we used the \textit{Ultimate Debian Database}\footnote{https://udd.debian.org/bugs/}, querying for known bugs for the packages installed in our container images. \textit{UDD} is a continuously updated system that gathers various \debian data in the same SQL database~\cite{nussbaum2010ultimate}: bugs, packages, upload history, maintainers, etc.

\udd contains information about all bug reports, including those that were archived. To identify if a package version is ``buggy", we queried \udd for all bug reports for that package. For each reported bug we checked if the specific package version was higher or equal to the version where the bug was first found. In case the bug report is resolved, we also verified if the package version is lower than the one fixing the bug.

\section{Empirical Analysis Results}
\label{sec:results}
\subsection*{$RQ_0$: How often are \docker images updated?}
\label{subsec:RQ0}

In order to analyze the technical lag of \docker container packages, it is essential to know how often \docker maintainers update their images and when they were last updated. This allows us to have a better understanding and carry out a fair comparison between the content of the different containers.

In September 2017, \textit{Anchore.io} analyzed the \official \docker images update history\footnote{https://anchore.com/blog/look-often-docker-images-updated/} and found that operating system images like \debian, \alpine or \textit{Ubuntu} update less often than non-OS images like \textit{Redis, MySQL} or \textit{Postgres}. They also noticed that \debian images are updated every month, which is the average compared to other OS images. Moreover, they observed that on some days many repositories push updates at the same time. Investigating this phenomenon, they found that in many cases this occurs the day after their base image \textit{debian:latest} was updated.

\begin{figure}[!ht]
	\begin{center}
		\setlength{\unitlength}{1pt}
		\footnotesize
		\includegraphics[width=1.0\columnwidth]{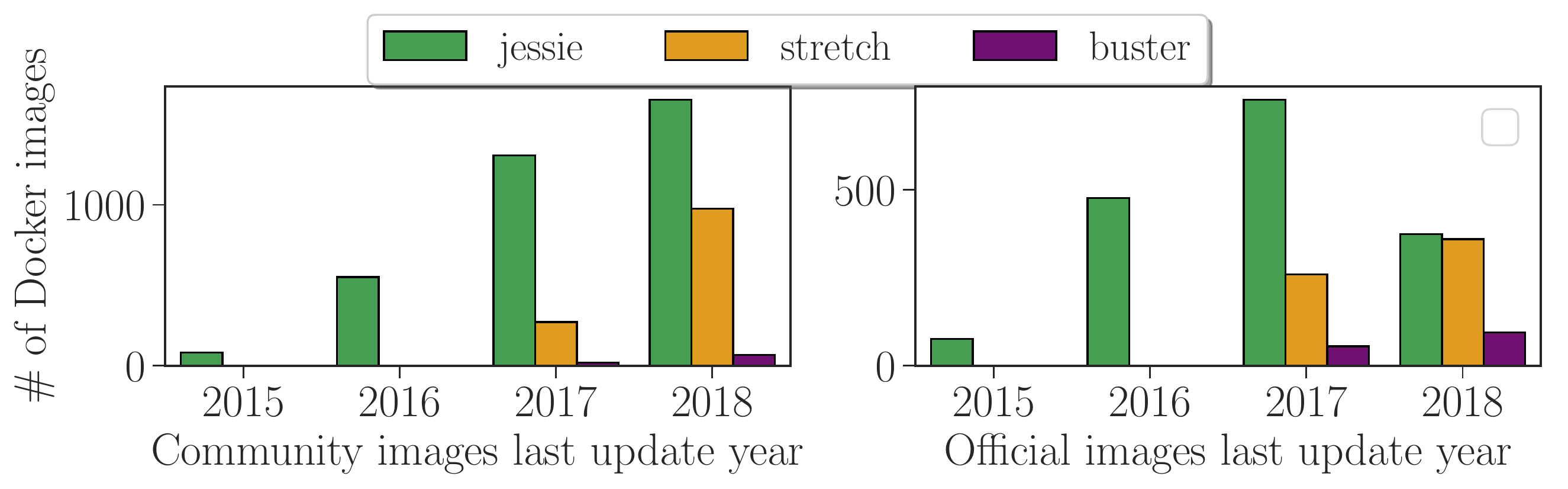}
		\caption{Year of \docker images last update, grouped by \debian distribution and community type (community or official)}
		\label{fig:last_updated}
	\end{center}
\end{figure}

\fig{fig:last_updated} shows the years when the considered \docker images were last updated. We observe that the number of official images that were updated in 2018 is less than those that were updated in 2017, for the images that make use of the \textit{Oldstable} version \jessie (\debian 8), while it is the opposite for the community images. Another important observation that should be taken into account for the rest of the study is that 48\% of the community images and 66\% of the official images were last updated before 2018. This can be explained by the number of images per repository: while official repositories have many images with different operating systems (\ie including slim and full) and for different architectures, community repositories have predominantly only one unique image (latest) with different tags. Thus, in official repositories new images emerge while others stop being updated; and community repositories tend to keep updating their images without creating new ones. 

\noindent\fbox{%
	\parbox{0.48\textwidth}{%
		\textbf{Findings for $RQ_0$}:  More than half of the \docker images have not been updated for four months.
	}%
}


\subsection*{$RQ_1$: What is the technical lag induced by outdated packages in containers?}
\label{subsec:RQ1}

\begin{figure}[!ht]
	\begin{center}
		\setlength{\unitlength}{1pt}
		\footnotesize
		\includegraphics[width=1.0\columnwidth]{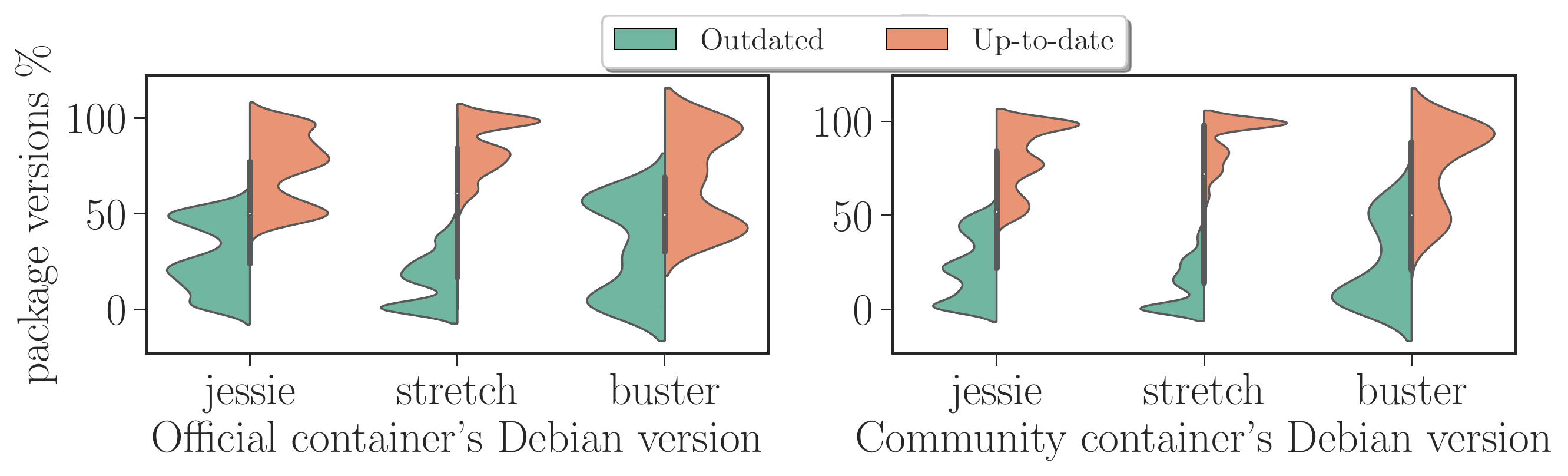}
		\caption{Proportion of up-to-date and outdated packages in \docker containers.}
		\label{fig:up_out_date}
	\end{center}
\end{figure}

$RQ_1$ investigates how outdated the packages in \docker containers are, based on a quantification of their technical lag.
Therefore, first we start by exploring how many packages within containers are up-to-date (\ie having the latest available fix). \fig{fig:up_out_date} shows the proportion of up-to-date and outdated packages in both \official and \community \docker containers, grouped by their \debian version. We observe that, regardless of the \debian version, most packages are up-to-date. The median proportion of up-to-date packages per container is 82\% of all installed packages. 
We also notice that packages inside \community containers are slightly more up-to-date (median 85\%) than packages inside \official containers (median 78\%) .

We statistically verified our observation by carrying out a non-parametric \textit{Mann-Whitney U} test that does not assume normality of the data. The null hypothesis assumes that the up-to-date package distributions of the \community and \official containers, grouped by their \debian version, are identical. For each pair of groups \textit{(Official-Jessie, Community-Jessie), (Official-Stretch, Community-Stretch)}, and \textit{(Official-Buster, Community-Buster)}, we rejected $H_0$ with statistical significance ($p<0.01$) when comparing the up-to-date package distributions of two groups of containers. However, for each comparison, we only found a small effect size ($|d| \leq 0.28$) using \textit{Cliff's Delta}, a non-parametric measure quantifying the difference between two groups of observations. 

When restricting our analysis to recent images only (\ie those that were last updated in 2018), we found that packages in \official containers are slightly more up-to-date than packages in \community containers.
Since \community images are based on \official images, this means that from all available \official images, \docker \community deployers tend to use the most recently updated ones, or they manually update all outdated packages inherited from old  \official images. We also found that the median proportion of up-to-date packages per container, in all cases, increased to 98\% of all installed packages. 

Once we have identified the outdated packages, we can calculate their technical lag.

We define \emph{technical lag} as the difference between the used package version and the latest available version. Since \debian maintainers are supporting three releases (\jessie, \stretchh and \buster), we measure the \emph{technical lag} in terms of the number of versions available between the used and the latest available version from the same \debian release. For instance, suppose that a used package \textit{q} of container \textit{c} has the following series of version numbers in \debian 8 \jessie (1.0.0, 1.1.1, 1.1.2, 1.1.4) and in \debian 9 \stretchh (2.0.0, 2.1.1). If we find that the used version of package \textit{q} is 1.1.1, then we compare it to the latest version of the available releases in \debian 8 \jessie, which is 1.1.4. The version lag in this case would be 2 versions, namely 1.1.2 and 1.1.4.  

\begin{figure}[!ht]
	\begin{center}
		\setlength{\unitlength}{1pt}
		\footnotesize
		\includegraphics[width=1.0\columnwidth]{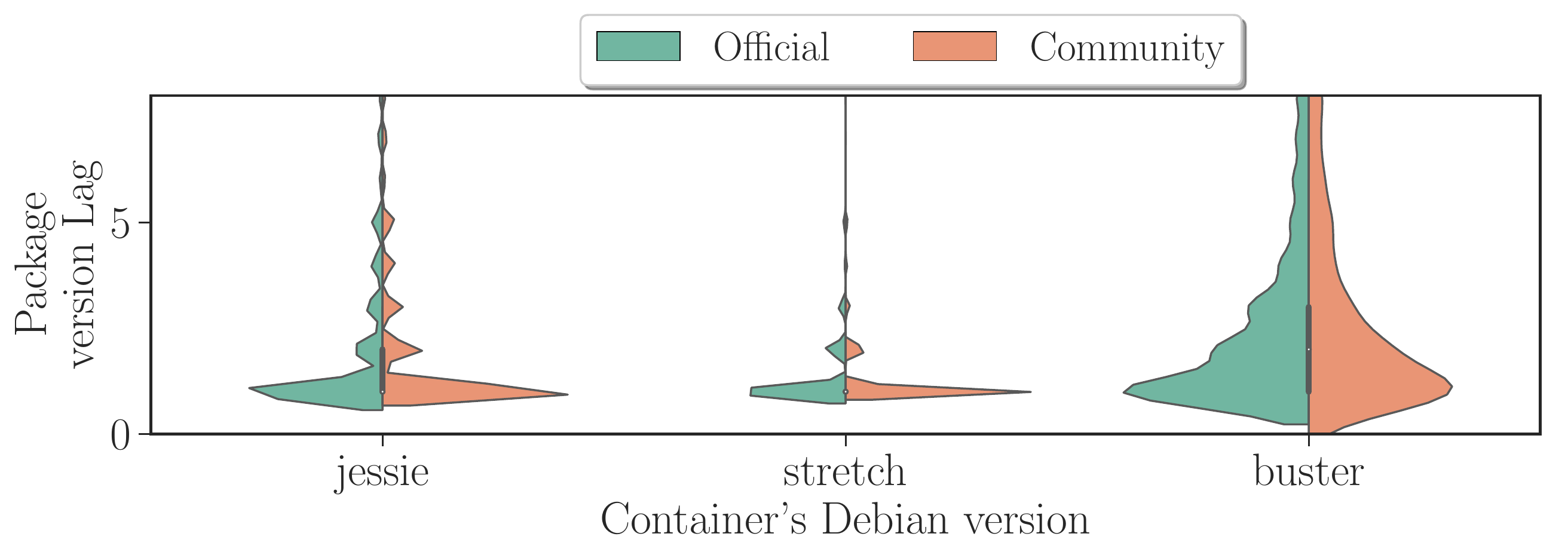}
		\caption{Version lag induced by outdated packages in \docker containers.}
		\label{fig:version_lag}
	\end{center}
\end{figure}

For all containers, we measured the technical lag of their outdated packages. \fig{fig:version_lag} shows the technical lag in terms of versions for the outdated packages in \docker containers, grouped by their \debian version. At first sight, we observe that the distributions are highly skewed. However, the distribution for the containers using \debian \stretchh is more highly skewed than the others. \tab{version_lag_tab} shows that the median version lag for both \jessie and \stretchh containers is 1, while it is 2 versions for \buster. This small difference is related to the state of the \debian release. Because the \buster version is now in the \textit{Testing} phase, many containers prefer not to depend on its packages since they are still subject to many changes, making it hard to keep up with its updating process. However, we can conclude that, in general, packages in \docker containers are either up-to-date or lagging behind with a median of 1 to 2 versions.

\begin{table}[!ht]
\centering
\caption{Mean and median of outdated package version lag grouped by \debian version and container type.}
\label{version_lag_tab}
\begin{tabular}{l|cc|cc}
\multirow{2}{*}{Containers} & \multicolumn{2}{c|}{Official} & \multicolumn{2}{c}{Community} \\ \cline{2-5} 
                            & mean               & median              & mean                & median               \\ \hline
\jessie                      & 2.13               & 1                   & 1.93                & 1                    \\ 
\stretchh                     & 1.36               & 1                   & 1.24                & 1                    \\ 
\buster                      & 2.71               & 2                   & 3.05                & 2                    
\end{tabular}
\end{table}

Since we found that the proportion of up-to-date installed packages per container is high, we decided to investigate more about the used package versions and when they were created. 
Considering only the up-to-date package versions this time, we traced back the date when they were first seen in \debian. 
We found that most of the package versions are old. Moreover, we found that exactly 80\% of the used \stretchh package versions and 90\% of the \jessie package versions were created before 2017-06-18, the release date of the \textit{Stable} version of \stretchh. We also found that 63\% of the used \jessie package versions were created before the release date of the \textit{Stable} version of \jessie (\ie 2015-04-25). This means that used packages tend to remain up-to-date because of the way in which \debian maintainers are creating and updating their packages.
%

\noindent\fbox{%
	\parbox{0.48\textwidth}{%
		\textbf{Findings for $RQ_1$}:
		\\ $\LargerCdot$ One out of five installed packages in containers is outdated.
		\\ $\LargerCdot$ Users of community containers tend to use recently updated official images.
		\\ $\LargerCdot$ Outdated installed packages are lagging behind one to two versions.
	}%
}


\subsection*{$RQ_2$: How vulnerable are packages in containers?}
\label{subsec:RQ2}

Verifying a software system for presence of security vulnerabilities is important. A vulnerability is a fault that could be exploited to abuse the system. With $RQ_2$, we analyze the vulnerability of \docker containers and if the observed vulnerability is related to the presence of outdated packages in the containers.

\begin{figure}[!ht]
	\begin{center}
		\setlength{\unitlength}{1pt}
		\footnotesize
		\includegraphics[width=1.0\columnwidth]{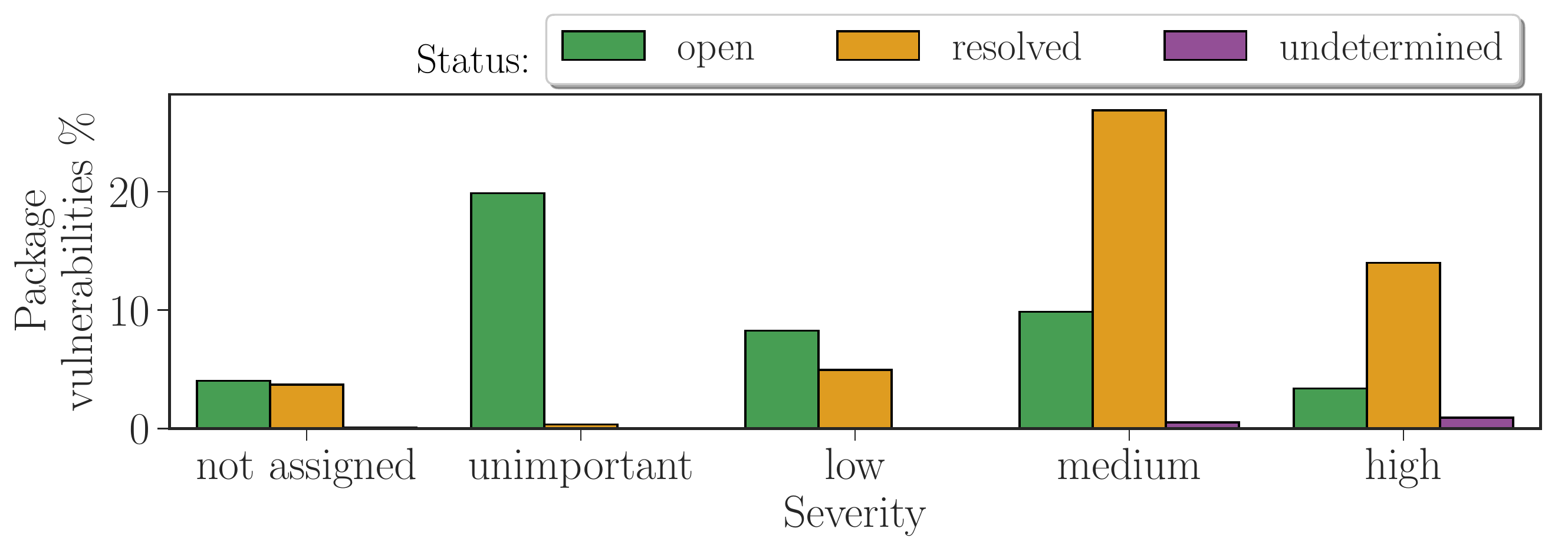}
		\caption{Proportion of found vulnerabilities grouped by their severity and status.}
		\label{fig:all_vuls}
	\end{center}
\end{figure}

We have therefore identified all vulnerable packages: only 12.2\% (\ie 488 out of 3,975) of all unique installed packages (from both \official and \community containers) had security issues. 
\fig{fig:all_vuls} shows the distribution of vulnerabilities by their severity (not assigned, unimportant, low, medium or high) and status (open, resolved and undetermined). We found that 49.9\% (\ie 12,806) of all vulnerabilities are \textit{resolved}, while 48.6\% (\ie 12,479) are still \textit{open}. A small proportion of 1.6\% (\ie 401) are \textit{undetermined}. We also observe that the majority of vulnerabilities has a \textit{medium} (37.2\%), \textit{unimportant} (20.2\%) or \textit{high} (18.3\%) severity. However, we found that all containers are affected by these severity vulnerabilities. In fact, 96\% of all containers are affected by all types of vulnerabilities, except for the \textit{not assigned} vulnerabilities. This possibly means that this small proportion of 12.2\% of packages cause the vulnerability of nearly all \docker containers. 

Next, we computed the number of vulnerabilities per container. We obtained a mean number of 1,336.44 vulnerabilities and a median of 601. The big difference between mean and median signals a heavily skewed distribution. 
Indeed, the maximum is 7,338 vulnerabilities for one container. 
Using a \textit{Mann-Whitney U} test we found statistically significant differences ($p<0.01$) in the number of vulnerabilities per container between \official and \community distributions. However, the effect size was small ($|d|<0.3$). \tab{container_vuls} shows more details about the distribution of the number of vulnerabilities in \docker containers.

\begin{table}[!ht]
\centering
\caption{Minimum, median and maximum number of vulnerabilities per container, grouped by \debian version and container type.}
\label{container_vuls}
\begin{tabular}{c|rrr|rrr}
\multirow{2}{*}{\begin{tabular}[c]{@{}c@{}}Containers\\ Debian\end{tabular}} & \multicolumn{3}{c|}{Official} & \multicolumn{3}{c}{Community}          \\ \cline{2-7} 
                                                                             & min     & median    & max     & min                     & median & max  \\ \hline
\jessie                                                                       & 155     & 658     & 7,106    & 155                     & 916  & 7,338 \\ 
\stretchh & 85      & 242     & 3,498    & 75                      & 336  & 4,729 \\ 
\buster & 34      & 134     & 659     & 41 & 183  & 1,035 
\end{tabular}
\end{table}

\begin{figure}[!ht]
	\begin{center}
		\setlength{\unitlength}{1pt}
		\footnotesize
		\includegraphics[width=1.0\columnwidth]{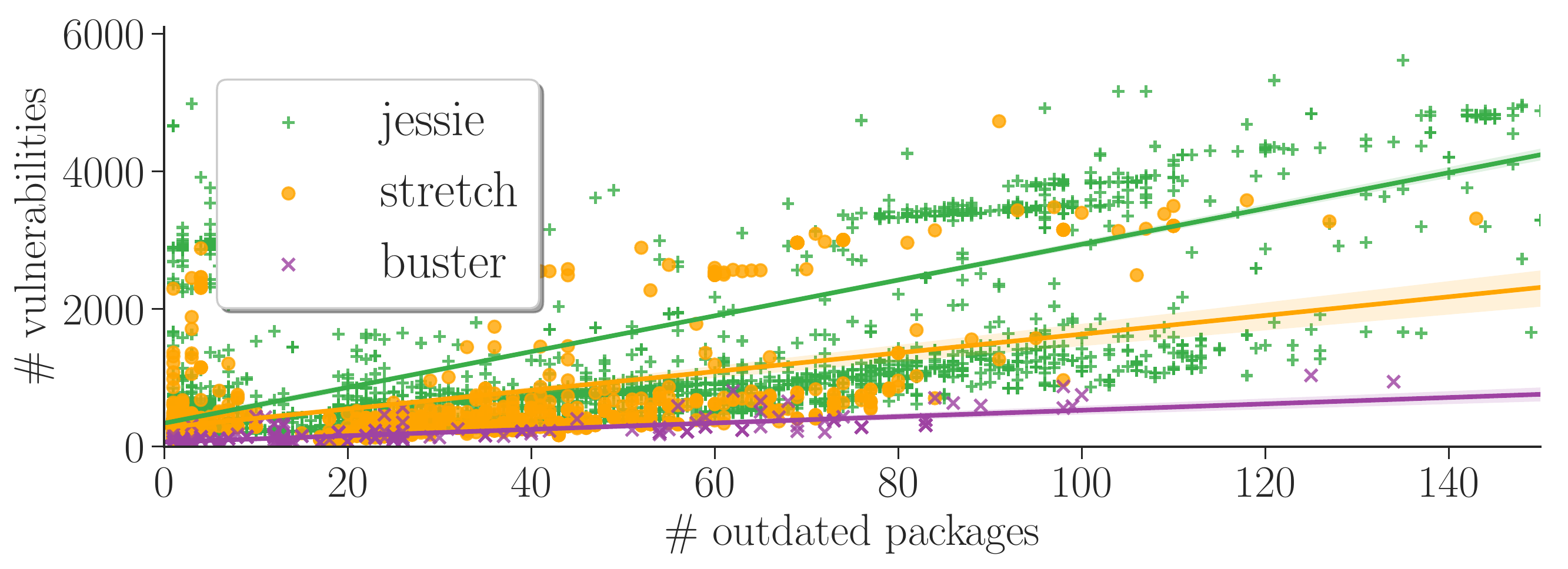}
		\caption{Number of outdated packages and vulnerabilities per container.}
		\label{fig:vuls_outdate}
	\end{center}
\end{figure}

To study the relation between the outdated container packages and the vulnerability of the container, we compared the number of outdated packages and number of vulnerabilities per container. Considering both \official and \community containers, and without differentiating between vulnerabilities by severity or status, we plot the numbers in a scatter plot  for different \debian versions (\fig{fig:vuls_outdate}). We visually observe a certain relationship between both metrics, especially for the \jessie containers: when the number of outdated packages increases, the number of vulnerabilities tends to increase as well. 

To verify our observations, we calculated Pearson's correlation coefficient $R$ and Spearman's $\rho$ over all packages, using the following threshold values 
: $0<|very\: weak|\leq 0.2 < |weak| \leq 0.4 < |moderate| \leq 0.6< |moderately\: strong| \leq 0.8 < |strong|\leq 1$.
A moderately strong increasing correlation (\textit{$0.6<R \leq 0.8$} and \textit{$0.6 < \rho \leq 0.8$}) for \jessie and \buster, and only a moderate one for \stretchh ($R=0.53$ and $\rho=0.42$) exists. 

%


\tab{top_vulnerable} shows the top 10 vulnerable \official and \community images. Their number of vulnerabilities and their age in months are given in parentheses. For the \official images, only the top vulnerable image in a repository is presented. For example, the \textit{perl} repository has many images with a high number of vulnerabilities; summed together, \textit{perl} would be in the top 10. However, since these images provide the same functionality, we report only one: the most vulnerable image. A common characteristic about the top 10 vulnerable containers for both types is that they have not been updated lately. They are between 10.7 months and three years old. 


\tab{package_vul_top} shows the top 10 most and least vulnerable source packages, with their number of vulnerabilities (in parentheses) and the proportion of containers that make use of them. The three most vulnerable source packages \textit{linux, chromium-browser} and \textit{imagemagick} seem to have high number of binary packages: 433, 419 and 327, respectively. We verified if there is a relation between the number of binary packages and the number of vulnerabilities, but we could not find a significant one. For instance, the source packages \textit{mono} and \textit{libreoffice} have 241 and 195 binary packages, but they have only 1 and 9 vulnerabilities, respectively.

\begin{table}[!ht]
\centering
\caption{Top 10 vulnerable official and community \docker images. Age is given in months.}
\label{top_vulnerable}
{\small
\begin{tabular}{p{2.7cm}|p{5cm}}
\begin{tabular}[c]{@{}l@{}}\textbf{Official images} \\  (\#vulnerabilities, age) \end{tabular}                                                              & \begin{tabular}[c]{@{}l@{}} \textbf{Community images}\\(\#vulnerabilities, age)   \end{tabular}                                                                                  \\ \hline
\begin{tabular}[c]{@{}l@{}} perl:5.12.5-threaded \\ (7106, 36.0) \end{tabular}  &   \begin{tabular}[c]{@{}l@{}} weboaks/chromium-xvfb-node\\ (7338, 27.8) \end{tabular}  \\ 
\begin{tabular}[c]{@{}l@{}} node:0.8.28 \\ (6889, 32.9) \end{tabular}  &  \begin{tabular}[c]{@{}l@{}} jmoifutu/almakioski-processor-base\\ (7282, 35.2) \end{tabular}  \\ 
\begin{tabular}[c]{@{}l@{}} erlang:18.2.1 \\ (6713, 28.4) \end{tabular}  &  \begin{tabular}[c]{@{}l@{}} suitupalex/node-composer \\ (7167, 26.2) \end{tabular}  \\ 
\begin{tabular}[c]{@{}l@{}} ruby:2.1.8 \\ (6426, 25.9) \end{tabular}  &  \begin{tabular}[c]{@{}l@{}} youdowell/php-fpm-for-wordpress\\ (7155, 25.5) \end{tabular}  \\ 
\begin{tabular}[c]{@{}l@{}} sentry:7.5.0 \\ (5742, 35.9) \end{tabular}  &  \begin{tabular}[c]{@{}l@{}} newsdev/github-keys \\  (7106, 35.5) \end{tabular}  \\ 
\begin{tabular}[c]{@{}l@{}} pypy:2-2.5.1 \\ (5742, 36.0) \end{tabular}  &  \begin{tabular}[c]{@{}l@{}} jmvrbanac/hastebin \\  (7106, 35.9) \end{tabular}  \\ 
\begin{tabular}[c]{@{}l@{}} python:2.7.9 \\ (5742, 36.2) \end{tabular}  &  \begin{tabular}[c]{@{}l@{}} devdetonator/protractor \\  (7106, 35.7) \end{tabular}  \\ 
\begin{tabular}[c]{@{}l@{}} gcc:6.1.0 \\ (5288, 20.5) \end{tabular}  &  \begin{tabular}[c]{@{}l@{}} kosmtik/kosmtik \\  (7106, 35.7) \end{tabular}  \\ 
\begin{tabular}[c]{@{}l@{}} hylang:0.12.1 \\ (3847, 11.1) \end{tabular}  &  \begin{tabular}[c]{@{}l@{}} misterbisson/couchbase-cloud-benchmark\\ (7106, 35.7) \end{tabular}  \\ 
\begin{tabular}[c]{@{}l@{}} elixir:1.4.4 \\ (3843, 10.7) \end{tabular}  &  \begin{tabular}[c]{@{}l@{}} pelle/ruby-phantomjs \\  (7013, 35.6) \end{tabular}  
\end{tabular}
}
\end{table}

\begin{table}[!ht]
\centering
\caption{Top 10 of most and least vulnerable \debian source packages.}
\label{package_vul_top}
{\small
\begin{tabular}{p{2.8cm}r|p{2.4cm}r}

\begin{tabular}[c]{@{}l@{}}\textbf{Most vulnerable}\\\textbf{source package}\\ (\# vulnerabilities) \end{tabular}          & \textbf{Used by} & \begin{tabular}[c]{@{}l@{}}\textbf{Least vulnerable}\\\textbf{source package}\\ (\# vulnerabilities) \end{tabular}    & \textbf{Used by} \\ \hline

linux (433)            & 54.51\% 

& audit (1) & 100.0\% \\ 
chromium-browser (419)     & 0.43\% 

& bzip2 (1) & 100.0\% \\ 
imagemagick (327)            & 28.13\%   

& db5.3 (1) & 100.0\% \\ 
php5 (186)          & 2.3\%

& pam (1) & 100.0\% \\ 
firefox-esr (139)         &  0.09\%

& sensible-utils (1) & 98.97\% \\ 
openjdk-7 (136) & 3.69\%

& libffi (1) & 94.13\% \\ 
tcpdump (132)            & 0.19\%

& cyrus-sasl2 (1) & 88.79\% \\ 
binutils (124)    & 53.55\%

& libssh2 (1) & 82.64\% \\ 
qemu (117)         & 0.22\%

& cryptsetup (1) & 73.43\% \\ 
mysql-5.5 (103)           & 26.82\%

& libbsd (1) & 72.8\% \\ 
\end{tabular}
}
\end{table}

\noindent\fbox{%
	\parbox{0.48\textwidth}{%
		\textbf{Findings for $RQ_2$}:
		\\ $\LargerCdot$ Nearly half of the vulnerabilities have no fix.
		 \\ $\LargerCdot$ All containers have high severity vulnerabilities.
		\\ $\LargerCdot$ The number of vulnerabilities depends on the \debian release used.
		\\ $\LargerCdot$ The number of vulnerabilities is moderately correlated with the number of outdated packages in a container.
	}%
}


\subsection*{$RQ_3$: To which extent do containers suffer from bugs in packages?}
\label{subsec:RQ3}


$RQ_3$ studies the presence of non-security-related bugs in \docker container packages, and the relation between bugs and outdated packages.
Considering all packages for both \community and \official images, we found that 50.1\% (1,994 out of 3,975) of all unique installed source packages have bugs. We also discovered that all containers have ``buggy'' packages.

\begin{figure}[!ht]
	\begin{center}
		\setlength{\unitlength}{1pt}
		\footnotesize
		\includegraphics[width=1.0\columnwidth]{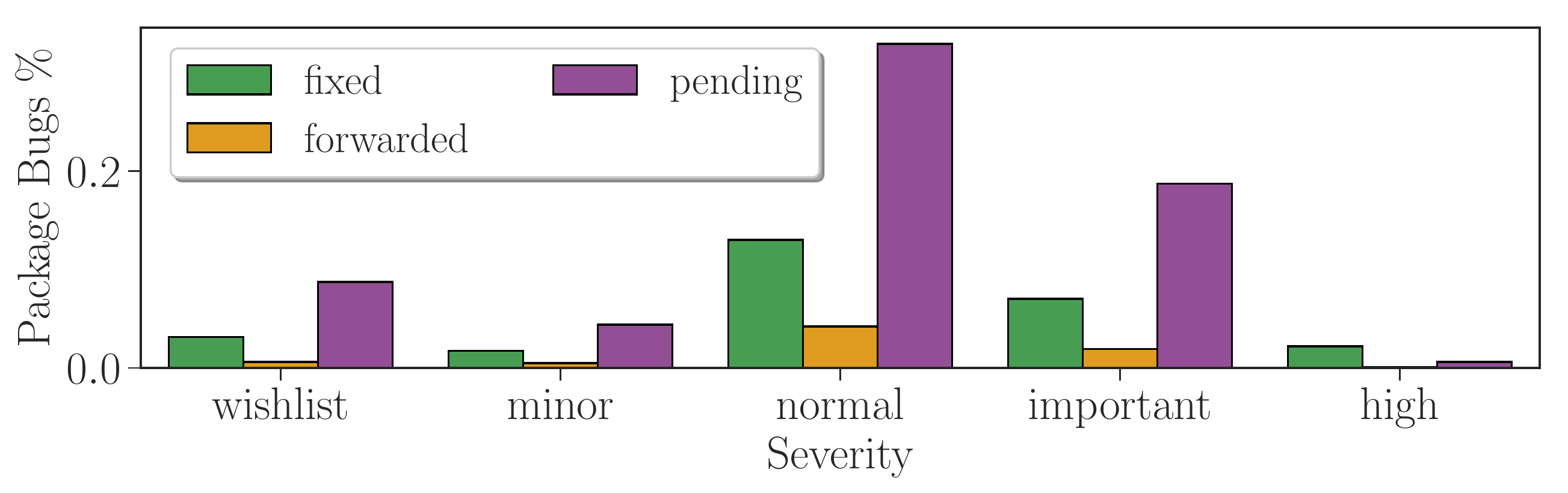}
		\caption{Proportion of bugs grouped by severity and status.}
		\label{fig:all_bugs}
	\end{center}
\end{figure}

\fig{fig:all_bugs} shows the distribution of bugs grouped by status (pending, forwarded, fixed) and severity (wishlist, minor, normal, important, high). The \textit{high} category combined three different severity types: \textit{serious, grave} and \textit{critical}.
We found that 65.5\% (12,863) of all bugs are still \textit{pending}, 7.3\% (3,460) are \textit{forwarded} and only 27.2\% (30,922) are \textit{fixed}. With respect to the severity, only 2.9\% of all bugs are \textit{high}, 27.7\% are \textit{important}, 50.2\% are normal and the rest is \textit{minor} or still in the \textit{wishlist}.


Since the majority of bugs are still \textit{pending} --nearly two out of three package bugs (65.5\%) are without a fix, and one out of two packages (\ie 50.1\%) has a bug--
we would expect the number of bugs per container to be higher than the number of vulnerabilities. 

We found that the mean and median numbers of bugs per container (including both \official and \community) are 2,081 
and 2,163 respectively. 
When comparing number of bugs in \official and \community containers, we obtained a statistically significant difference with $p<0.01$ using the \textit{Mann-Whitney U} test. However, a small effect size ($|d|<0.2$) for the \buster and \stretchh versions, and a medium effect size ($|d|=0.28$) for the \jessie version exist. The number of reported bugs decreases with more recent versions of \debian.
\tab{container_bugs} shows details about the distribution of the number of bugs in \docker containers.

\begin{table}[!ht]
\centering
\caption{Min, median and max of bugs per container grouped by \debian version and container type.}
\label{container_bugs}
\begin{tabular}{c|rrr|rrr}
\multirow{2}{*}{\begin{tabular}[c]{@{}c@{}}Containers\\ Debian\end{tabular}} & \multicolumn{3}{c}{Official} & \multicolumn{3}{c}{Community}            \\ \cline{2-7} 
                                                                             & min     & median    & max     & min                      & median & max   \\ \hline
\jessie                                                                       & 1,307    & 2,201     & 3,415   & 1,296                     & 2,450  & 5,628 \\ 
\stretchh                                                                      & 962     & 1,683     & 2,665   & 828                      & 1,759  & 3,285 \\ 
\buster                                                                       & 213     & 560       & 776     & 278 & 561    & 1,098 
\end{tabular}
\end{table}

\begin{figure}[!ht]
	\begin{center}
		\setlength{\unitlength}{1pt}
		\footnotesize
		\includegraphics[width=1.0\columnwidth]{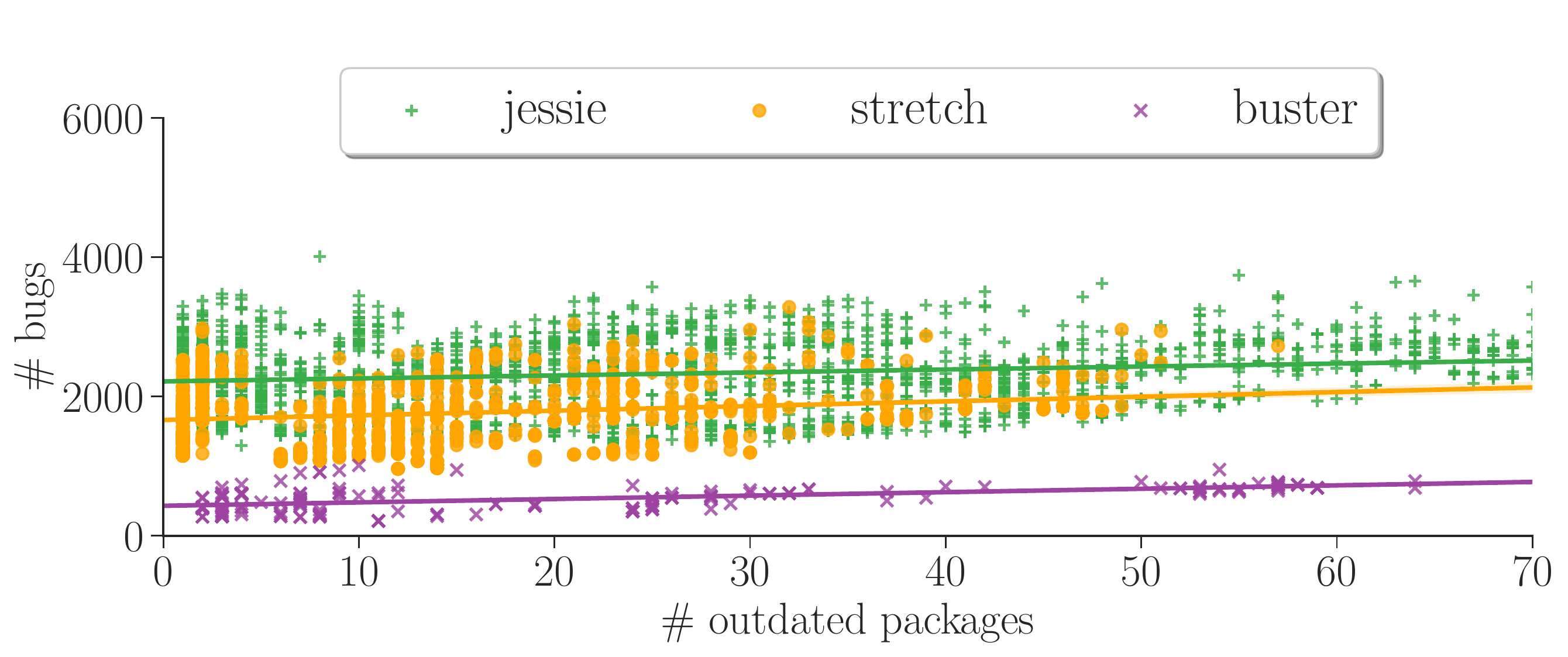}
		\caption{Number of outdated packages and bugs per container.}
		\label{fig:bugs_outdate}
	\end{center}
\end{figure}

We also studied the relation between the presence of outdated packages and bugs in containers. 
Considering both \official and \community containers, and without differentiating between bug status or severity, \fig{fig:bugs_outdate} shows a scatter plot, for different \debian versions, of the relation between the number of outdated packages and the number of bugs found in each container. Opposite to what we observed for vulnerabilities in $RQ_2$ (\fig{fig:vuls_outdate}), there is only a relation between the number of bugs and number of outdated packages for the \buster version.

To statistically verify our observations, we computed Pearson's $R$ and Spearman's $\rho$ correlation coefficients for all packages. For \jessie and \stretchh, the correlation was weak to very weak ($R\leq0.2$ and $\rho\leq0.22$). For \buster, a moderate increasing correlation ($R=0.58$ and $\rho=0.55$) was observed. 
This is because \buster contains many new package versions, which still get new bug notifications.
\jessie and \stretchh, on the other hand, contain mainly stable and old package versions, that mainly contain old bugs. Indeed, 90\% of \jessie bugs were created before July 2016, and 78\% of \stretchh bugs were created before its \textit{Stable} release in June 2017. 

\smallskip
\noindent\fbox{%
	\parbox{0.48\textwidth}{%
		\textbf{Findings for $RQ_3$}:
		 \\ $\LargerCdot$ All containers have buggy packages.
		\\ $\LargerCdot$  65\% of bugs in installed packages are without a fix.
		\\ $\LargerCdot$ The number of bugs is related to the \debian version used.
		\\ $\LargerCdot$ There is a weak correlation between the number of bugs and the number of outdated packages in containers relying on the \textit{Stable} and \textit{Oldstable} \debian release.
	}%
}


\subsection*{$RQ_4$: How long do bugs remain unfixed?}
\label{subsec:RQ4}

Since nearly half of all vulnerabilities are still \textit{open} and 65\% of all bugs are still \textit{pending}, $RQ_4$ investigates how long it takes for a bug to get fixed.
To do so, for all bugs, we compute the time interval between the bug report creation date in the \udd and the last modification date of the bug, considering that this corresponds to the bug fix date, in case a fix was observed.

\begin{figure}[!ht]
	\begin{center}
		\setlength{\unitlength}{1pt}
		\footnotesize
		\includegraphics[width=1.0\columnwidth]{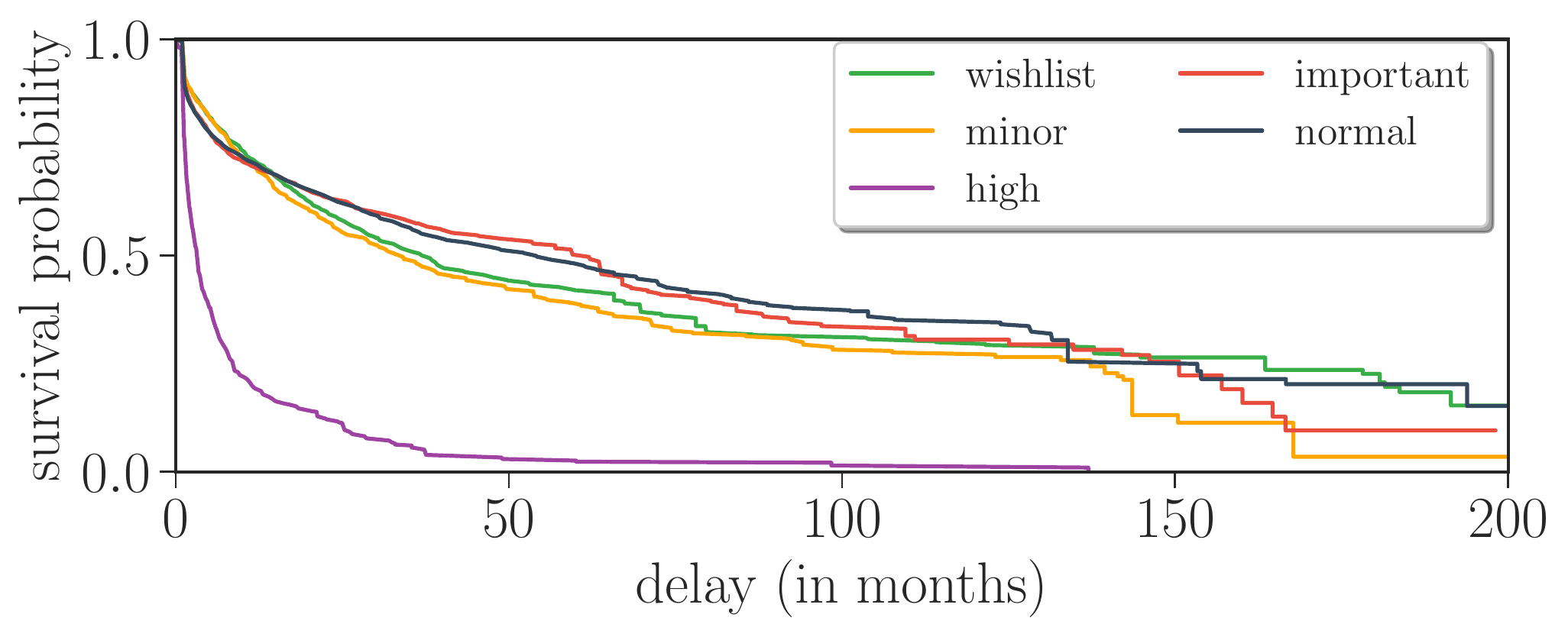}
		\caption{Survival probability for event ``bug is fixed" w.r.t. the bug report creation date.} 
		\label{fig:survival_bugs}
	\end{center}
\end{figure}

We rely on the statistical technique of survival analysis based on the non-parametric \textit{Kaplan-Meier} statistic estimator commonly used to estimate survival functions\cite{goel2010understanding} (widely used before in software engineering research~\cite{samoladas2010survival,scanniello2011source,lin2017developer}).
\fig{fig:survival_bugs} shows survival curves per severity level for the event ``bug is fixed" w.r.t. the bug report creation date. We observe that the time to fix a bug does not always depend on its severity level. \textit{High} severity bugs are fixed faster than other kind of bugs. 
For example, it takes 53.8 and 33.5 months so that 50\% of all \textit{normal} and \textit{minor} bugs get fixed, respectively, while it only takes  3 months for \textit{high} severity bugs.  
\debian maintainers prefer to start with easy bugs that are trivial to fix rather than \text{normal} ones. Nonetheless, bugs that may have an impact on releasing the package with the \textit{Stable} release of \debian (\ie \textit{high severity}\footnote{\url{https://www.debian.org/Bugs/Developer.en.html\#severities}}) have the highest priority.

To find out if there are statistically significant differences between the survival curves per severity, we carried out log-rank tests for each severity pair. The differences were statistically confirmed ($p<0.01$ after Bonferroni correction) 
except for the pairs \textit{(normal, important)} and \textit{(minor, wishlist)} where the null hypothesis could not be rejected.



\smallskip
\noindent\fbox{%
	\parbox{0.48\textwidth}{%
		\textbf{Findings for $RQ_4$}: 
	 \\ $\LargerCdot$ Normal and minor bugs require in the median very long times to be fixed (53.8 and 33.5 months). 
	 \\ $\LargerCdot$ High severity bugs are fixed ten times faster than other kind of bugs.}%
}
\subsection*{$RQ_5$: How long do security vulnerabilities remain unfixed?}
\label{subsec:RQ5}

\begin{figure}[!ht]
	\begin{center}
		\setlength{\unitlength}{1pt}
		\footnotesize
		\includegraphics[width=0.9\columnwidth]{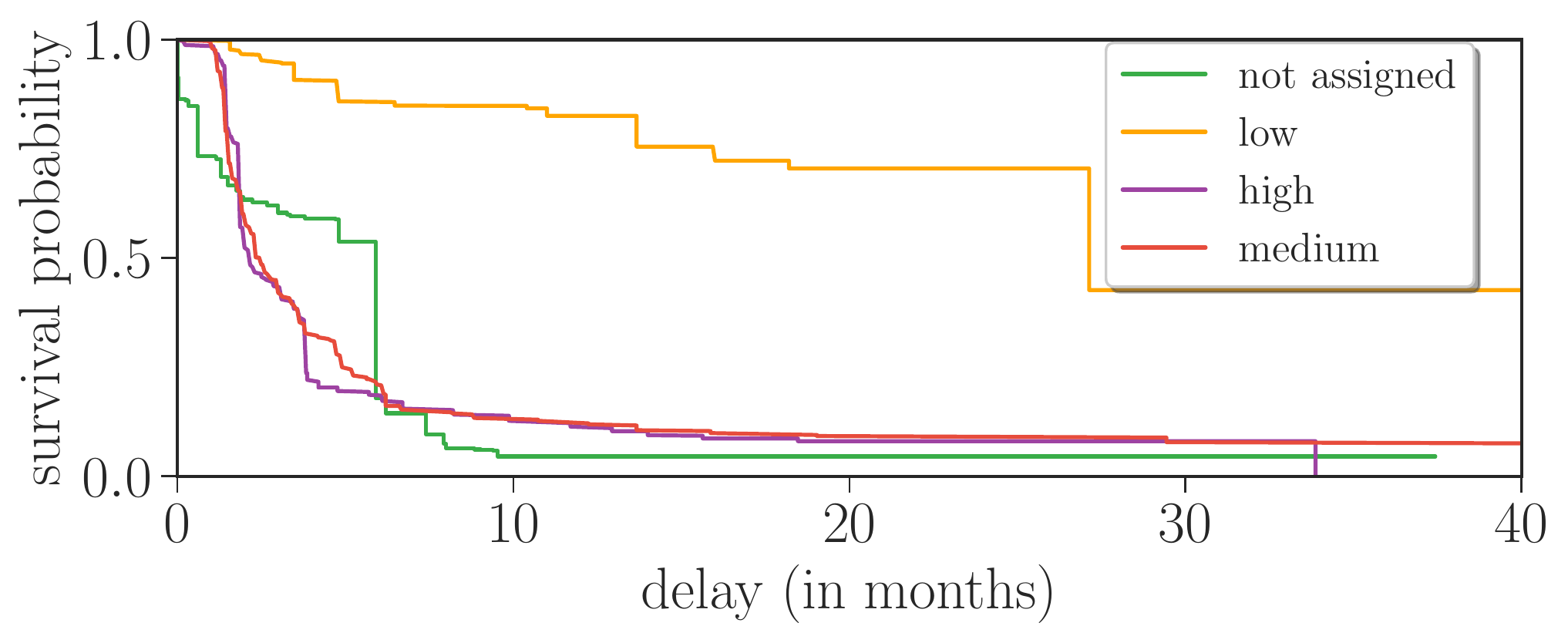}
		\caption{Survival probability per severity level for event "security vulnerability is fixed" w.r.t. the date of the bug arrival.} 
		\label{fig:survival_vuls}
	\end{center}
\end{figure}

Similar to the bug survival analysis, we analyzed the survival of security vulnerabilities over time. Using the \debian security tracker we extracted the \textit{debianbug id} for each vulnerability. With this \textit{id}, we searched in the \udd for the creation and last modification date of the corresponding bug.
We only found 62\% of all vulnerabilities with a corresponding \textit{debianbug id}. This proportion of vulnerabilities is responsible for 93\% of all container vulnerabilities. For this subset we carried out a survival analysis for the event ``security vulnerability is fixed" w.r.t. the bug report creation date. \fig{fig:survival_vuls} shows the \textit{Kaplan-Meier} survival curves for each severity level found on the security tracker (as opposed to the severity of the bug reported in the \udd).
 
We observe that vulnerabilities are fixed faster than other types of bugs. It takes 5.9 months for \textit{not assigned} severity vulnerabilities, 2.4 months for \textit{medium} severity vulnerabilities, and 2.1 months for \textit{high} severity vulnerabilities. \textit{Low} severity vulnerabilities take much more time to fix: 27 months to fix 50\% of them. We could not include the \textit{unimportant} vulnerabilities since only 0.5\% of them are fixed. We carried out log-rank tests to compare whether there are statistically significant differences between the survival curves depending on the severity of the bug. We could reject the null hypothesis assuming the similarity between the survival analysis curves with statistical significance ($p<0.01$ after Bonferroni correction), except for \textit{high} and \textit{medium} severity vulnerabilities.

The findings concerning the \textit{unimportant} vulnerabilities could be understandable since those vulnerabilities do not affect the binary package, but only materials and files that are not built (\eg doc/foo/examples/\footnote{https://security-team.debian.org/security\_tracker.html}). 
To investigate this further, we identified the bug severity of the \textit{low} severity vulnerabilities inside the \debian bug tracker (\ie from the \udd), and found that 77.5\% of these vulnerabilities have an \textit{important} or \textit{normal} bug severity. This correlates with the findings in $RQ_4$ and explains the results for the \textit{low} vulnerabilities. However, we also observed that 51\% of the \textit{high} severity vulnerabilities are labeled as \textit{important} bugs and 45\% of them are considered as \textit{high} (\ie \textit{serious, grave} or \textit{critical}) bugs. This means that an upstream high vulnerable package can have a different priority downstream, depending on the downstream maintainers assessment (\ie how a package is compiled, how it is integrated into the distribution, etc.).




\smallskip
\noindent\fbox{%
	\parbox{0.48\textwidth}{%
		\textbf{Findings for $RQ_5$}:
		 \\ $\LargerCdot$ \textit{High} and \textit{medium} severity vulnerabilities are fixed faster than \textit{low} severity vulnerabilities.
		 \\ $\LargerCdot$ Vulnerability reports \emph{upstream} might have different severity \emph{downstream}.
	}%
}

\section{Discussion and Actionable results}
\label{sec:discussion}

In an ideal world, containers should depend on the most recent available version of their used packages, in order to benefit from the latest functionality, security updates and bug fixes. However, maintainers might be more focused on other software characteristics such as package stability, or they just choose not to upgrade certain packages because of the considerable effort that may be involved in doing so (``if it ain't broke, don't fix it''). For this reason, we studied the presence of technical lag in \docker containers, and related it to the presence of bugs and severity vulnerabilities. 

In~$RQ_1$, we found that, in general, \debian packages used in \textit{stable} releases are old and up-to-date (\ie having the latest available fix). This implies that it should be easy for developers to keep up with the \debian updating process, since package maintainers are not releasing often. Moreover, new package versions in the \textit{Stable} and \textit{Oldstable} releases are only about security patches, so there is little to fear of breaking changes. For the \textit{Testing} release, however, things are different: deployers should be aware about how the \debian project works.

\smallskip
\noindent\fbox{%
	\parbox{0.47\textwidth}{%
		\textbf{Actionable result}: Container deployers should be aware that the optimal update frequency of their base images and installed packages depends on the base \debian version.
	}%
	}
	
\vspace{0.2cm}

In~$RQ_2$, we found that the number of vulnerabilities is related to the number of outdated packages per \debian-based container. This demonstrates that containers could benefit from better updating procedures, allowing them to avoid security issues coming from their installed packages. Moreover, we found that the number of vulnerabilities is related to the \debian release. 

\smallskip
\noindent\fbox{%
	\parbox{0.47\textwidth}{%
		\textbf{Actionable result}: Deployers who prefer stability to new functionalities are recommended to use the Stable and Oldstable versions that include only the most important corrections or security updates. 
		To have a lower number of severe vulnerabilities, container deployers using the \textit{Oldstable} \debian release should upgrade to the \textit{Stable} release.
		
	}%
	}

\vspace{0.2cm}

In~$RQ_3$, we found that all \docker containers are vulnerable and contain packages with a high number of vulnerabilities and bugs. Since we did not discover a high version lag in containers, we do not think it is the responsibility of \docker deployers to avoid all vulnerabilities. Even containers with up-to-date packages still may have a high number of vulnerabilities. 

\smallskip
\noindent\fbox{%
	\parbox{0.47\textwidth}{%
		\textbf{Lesson learned}: No release is devoid of vulnerabilities, so deployers cannot avoid them even if they deploy the most recent packages.
	}%
	}
	
\vspace{0.2cm}

 We could not find a significant relation between the number of outdated packages and the number of bugs. However, we observed that the number of bugs is related primarily to the \debian release. This means that deployers that care about bugs and new functionality and not about stability, should definitely upgrade to the \debian \textit{Testing} release, since the updates in \textit{Stable} and \textit{Oldstable} releases are primarily about security bugs (\ie severity vulnerabilities).

\smallskip
\noindent\fbox{%
	\parbox{0.47\textwidth}{%
		\textbf{Actionable result}: Container deployers concerned with having as few non-security bugs as possible should upgrade to the \textit{Testing} release, at the expense of having a lower package stability.
	}%
	}
\vspace{0.2cm}

Based on a survival analysis, we concluded that security vulnerabilities take less time to fix than other kind of bugs. The relation between vulnerabilities, bugs and outdated packages shows that container deployers should give a high priority to updating when checking their container packages.

\smallskip
\noindent\fbox{%
	\parbox{0.47\textwidth}{%
		\textbf{Actionable result}: High security bugs are first priority for \debian maintainers; they are fixed faster than other kind of bugs. Container deployers should be aware of the newly available versions of their installed packages and keep technical lag to the minimum to avoid this type of bugs.
	}%
	}
	
\vspace{0.2cm}

Comparing our results about vulnerabilities to previous observations~\cite{shu2017study}, we found \debian-based \docker containers to have an average number of vulnerabilities (\ie 460) that is above the average for all \docker containers (\ie 120). However, the number of vulnerabilities depends on the number of installed packages found. For example, it is not fair to compare vulnerabilities between \debian containers and \alpine\footnote{\alpine is a minimal
image based on the security-oriented, lightweight Alpine Linux distribution with a complete package index that is no more than 8 Mb in size.} containers, unless we compare their size as well (in terms of number of installed packages).

As highlighted before, it is important to verify not only vulnerabilities, but also other bugs. 
Indeed, bugs make the system behave in unexpected ways, resulting in faults, wrong functionality or reduced performance. Researchers already found that performance bugs are similar to security bugs, in that they require more experienced developers to fix them~\cite{zaman2011security}. Hence, it is essential to include bug analysis tools into existing automated scan and security management services such as \textit{Anchore.io} or \textit{Quay.io}. 

Moreover, an \textit{Anchore.io} survey showed that container deployers care more about package vulnerabilities than having packages up-to-date. However, we found that less outdated containers have less vulnerabilities. Thus, we believe that including the technical lag as a measure of how outdated packages are, can empower automated scan and security management tools to give better insights about the security of \docker containers.

\smallskip\noindent\fbox{%
	\parbox{0.47\textwidth}{%
		\textbf{Actionable result}: \docker scan and security management tools should improve their platforms by adding data about other kind of bugs and include the measurement of technical lag to offer deployers information of when to update.
	}%
	}
	
\vspace{0.2cm}

Using our method, and more specifically our automated data extraction tool, container deployers can check the state of their container \debian package vulnerabilities and bugs. They can also compare with other containers that make use of the \debian operating system.

\section{Threats to validity}
\label{sec:threats}

Our study was focused on \docker containers that make use of the Testing, Stable or Oldstable versions of \debian. 
The results of our analysis can therefore not be generalized to other base images in \docker. The analysis itself, however, can be easily replicated on other base images. 

We chose to use the technical lag as a measurement. We only compared the used package version with latest available version of the package within the same \debian release. Our results may differ when comparing with the latest available package versions from the latest (\eg \textit{Stable}, \textit{Testing} or \textit{Unstable}) \debian releases.

Moreover, it is not trivial to identify which package versions are affected by bugs or severity vulnerabilities. For example, the way in which we computed vulnerabilities and bugs was different. For vulnerabilities we relied only on the fixed version, since this is the way it is done in companies such as \textit{CoreOS} or \textit{Anchore.io}. For bugs, we relied on two sources of information: the bug report creation date and its last modification date. Counting bugs in the same way as vulnerabilities would result in more bugs than the ones considered in this analysis. 

Also, when searching for vulnerabilities in the Debian security tracker, the \textit{debianbug id} was not found for 38\% of the vulnerability reports. This may have influenced our survival analysis results. However, the missing proportion of vulnerability reports is responsible for only 7\% of all analyzed container vulnerabilities.

\section{Conclusions and future work}
\label{sec:conclusion}

This paper presented an empirical analysis of the state of packages in public \docker containers that are based on the \linux-based \debian distribution. We studied how outdated container packages are and how this relates to the presence of bugs and severity vulnerabilities.

Considering both \official and \community images, we studied 7,380 popular unique images. 
We observed that most container packages have the latest fix available in \debian, even for old packages (\eg \textit{Stable}). However, we found that all containers have vulnerable and buggy packages. Studying outdated packages in more detail, we found that their number is correlated with the number of vulnerabilities found in a container.

We observed that in \debian, taking care of security vulnerabilities is more important than taking care of bugs. This results a high number of open bugs for the \textit{Stable} and \textit{Oldstable} releases. Therefore, even up-to-date installed package versions could be affected by these open bugs.

These findings indicate that container deployers whose major concerns are stability and security need to rely on better updating procedures. In contrast, container deployers that care more about functionality and bugs should rely on the newest \debian releases.

When studying how outdated \docker images are, we did not differentiate between specific package characteristics such as their size, service, targeted audience, or provided functionality. Moreover, we did not differentiate between release types (\eg patch, minor or major) when calculating technical lag.
In future work we would like to consider other measures of technical lag while considering package characteristics and all available releases in a project. For instance, in many cases vulnerability fixes are first done in the \textit{Testing} or \textit{Unstable} releases before entering the \textit{Stable} and \textit{Oldstable} releases.

Besides the operating system packages, containers have other types of packages installed on them, for instance, \textit{PyPI} and \textit{npm} packages. Such packages can be vulnerable as well~\cite{decan2018impact,Lauinger2017}. Thus, we aim to include these other types of packages. We also plan to carry out a comparison with other operating systems and other base images.

Since our data extraction and analysis are automated, we aim to create a tool that automatically gathers and analyses package information, such as package vulnerabilities, bugs, possible updates, possible conflicts, dependencies,  \textit{etc.}, for different operating systems. To be able to analyze their evolution, we also aim to gather monthly snapshots about the \docker containers content and state. Finally, we want to carry out surveys and interviews with container deployers to validate the implication of our work and collect more insights about common practices.

\providecommand{\noopsort}[1]{}

\end{document}